\documentstyle[epsfig]{aipproc}
\begin{document}
\title{Current Issues}
\author{R. D. Blandford}
\address{130-33 Caltech\\Pasadena\\CA 91125}
\maketitle
\begin{abstract}
Cosmic explosions are observed in many astrophysical
environments. They range in scale from hydromagnetic instabilities
in the terrestrial magnetotail and solar ``nanoflares'' to cosmological 
gamma ray bursts, supernovae and the protracted
intervals of nuclear activity that produce the giant quasars and radio galaxies. There
are many parallels in the analyses of the explosion sites that are highlighted at
this workshop, specifically 
stellar coronae, accretion disks, supernovae and compact objects. 
In this introductory talk, some 
general issues are discussed and some more specific questions 
relating to the individual sites are raised.
\end{abstract}
\section*{Introduction}
A star lives its life, from its birth out of a loose
assembly of molecular gas till the time when it makes its quietus as a white dwarf,
neutron star or black hole, fighting gravity. Although it eventually 
looses the fight (unless it manages to make a Type Ia supernova),
it does not concede graciously. Time and again it finds itself
transitioning from a metastable equilibium to a state of lower energy
on a dynamical timescale. Similar principles govern the evolution
of galaxies where there can be runaway formation of massive stars
or episodic accretion onto the central, massive black hole. These 
are ``Cosmic Explosions'' - the topic of this workshop.

When I was first asked to introduce ``Current Issues'', I thought
the title curiously apposite because it is ultimately currents
- electrical, weak (charged and neutral) and (with some license) strong - 
that are responsible for these impulsive releases of energy. 
Out of the many astrophysical sites that could have been 
included on the program, the organizers have chosen to concentrate
on a few of the most interesting ones, that I shall consider in turn
- the solar corona, accretion disks surrounding
young stellar objects, novae, supernovae, ``hypernovae'' and jets.
As I am neither competent nor patient enough to describe these 
in any detail, I have chosen to list some recent advances in our
observational and theoretical understanding in each case and to pose 
a few questions some of which may already have answers which I hope 
subsequent speakers will provide.
In view of the large range of topics reviewed I cannot hope to give a
representative or even a useful bibliography, and so I shall 
give none and defer to subsequent speakers.
\begin{table}[b!]
\caption{{\bf Observed} Characteristics of Some Cosmic Explosions}
\label{table1}
\begin{tabular}{lddd}
Explosion&Energy (erg)&Timescale (s)&Power (erg s$^{-1}$)\\
\tableline
Solar Flares&$10^{32}$&$10^4$&$10^{28}$\\
FUORs&$10^{45}$&$10^9$&$10^{36}$\\
Novae&$10^{44}$&$10^6$&$10^{38}$\\
Supernovae&$10^{50}$&$10^6$&$10^{44}$\\
Hypernovae (GRBs)&$10^{53}$&$10^2$&$10^{51}$\\
Jets&$10^{61}$&$10^{14}$&$10^{47}$\\
\end{tabular}
\end{table}
\section*{Solar and Stellar Flares}
The combined observations of the YOHKOH, SOHO and TRACE satellites
are transforming our view of the solar corona and, consequently, of the surface activity of other 
stars. In particular they have given us an appreciation of the dynamics
of magnetic field 
lines as they are gently shuffled by underlying convective motions. The whole 
region above the photosphere is permeated by a "magnetic carpet" which is re-woven
every couple of days. The solar prominences and coronal arches, prominent in X-ray 
images are just the regions where the plasma happens to be hottest and, contrary to what
might have been thought, the magnetic field is {\it weakest}. This magnetic activity 
is intrinsically dissipative and this keeps the corona at million degree temperatures 
and launches the solar wind.   

The quiet solar wind appears
to be a simple and quasi-steady flow at least at high latititude (as measured by the Ulysses 
spacecraft) with poloidal and toroidal
magnetic field components declining as $\propto r^{-2},r^{-1}$, respectively. By contrast,
the equatorial 
outflow appears to be dominated by unsteady coronal mass ejections. The equatorial current sheet
is naturally unstable and develops its characteristic ``ballerina skirt'' sector structure.
(Perhaps something similar has been observed by Chandra in the Crab Nebula.)
Solar physics has much to teach us about accretion disks, where the underlying motions
are much faster and, necessarily, supersonic. It should be no surprize that they
are often accompanied by hot coronae, that dissipate a large fraction of the gravitational
energy release, and powerful outflows.

My list of questions includes:
\begin{itemize}
\item What are the true laws of astrophysical MHD? Traditional, global 
MHD has been based upon analytic solutions of the 
equations of conservation of mass, momentum and flux under conditions of high symmetry ignoring
dissipation. However, real MHD is heavily influenced
by the microphysical behavior of current sheets,
tiny reconnecting regions, shock fronts etc in much the same way that
hydrodynamic flows are beholden to boundary layers.  
Perhaps there are simple, phenomenological
rules which can reconcile these two approaches.
\item What is a solar flare? We know of many 
examples of magnetostatic configurations that can be slowly 
altered until they become unstable and release a large amount of magnetic energy.
However we do not understand which of these are most likely to occur in practice
and what is the partition of the release of energy between local heating and 
the bulk kinetic energy that drives
outgoing shock waves. (Similar questions exist in earthquake studies.)
\item What is the structure of shock fronts?
Simulation and {\it in situ} measurement has greatly improved our
understanding of collisionless shocks. It appears that thermal electrons are commonly 
transmitted with sub-equipartition energies, as is also found to be the case
with supernova shock waves. The 
detailed plasma physics still eludes us, though. This issue is related
to the question of the injection of suprathermal ions into the first order Fermi
acceleration process that appears to be responsible for producing
most Galactic cosmic rays.
\item What determines the energy and length scales that dominate coronal heating? 
The form of this dissipation appears to be primarily 
reconnection and to be dominated by frequent ``nanoflares'', (although 
this conclusion is still
controversial). This realization has, in turn, stimulated analysis of new
modes of magnetic reconnection.
Still, it is the occasional giant flare that commands our observational
attention and provides the most detailed diagnostics.
\item How is the solar wind launched? The observed coronal 
temperature is insufficient to give the gas its 700-800 km s$^{-1}$
outflow speed as measured by Ulysses.
This leaves hydromagnetic wave acceleration 
as the prime suspect. Understanding the acceleration and 
stability of the solar wind is highly relevant to the study of jets.
\end{itemize}
\section*{Young Stellar Objects}
Accretion disks and bipolar outflows appear to be a standard feature of 
star formation. The optical jets can propagate through the interstellar
medium over distances more than $\sim10$~pc, quickly polluting it with magnetic 
field and metals (an observation of some cosmological importance). 
However, this outflow is not steady. In particular, thermal instability of the accretion disk 
produces ``FU Ori'' outbursts where perhaps $\sim0.01$~M$_\odot$ of gas are expelled
with comparatively high speed over a decade or so every ten thousand years.
Smaller scale explosions create the ``Herbig-Haro'' objects  which are
presumably traveling forward-reverse shocks. These are sometimes observed
in matched pairs, one in each jet confirming that they originate at the disk.
The morphological similarity to, for example, the knots in the M87 jet
is clear.

Some important questions include:
\begin{itemize}
\item How much of the mass in the original protostellar disk accretes
onto the central protostar and how much is lost in the form of a wind or jet?
A related question is how much of the angular momentum is removed
in this manner as opposed to being transported radially in the disk to large
radii where it can, supposedly be extracted by large tidal torques.
\item How much mass and energy is associated with the FUOR outbursts and 
how much is transported in the long intervals between outbursts?  (It is not 
clear how far one can push the dynamical analogy but similar questions 
have been raised in trying to understand the Galactic microquasar GRS 1915+105.)
\item How are the optical jets collimated and confined? Magnetic collimation is
commonly invoked, but even here, several alternatives have been discussed. 
The field may be primarily vertical near the disk and shape the outflow
through magnetic pressure. Alternatively, the field may have a significant 
radial component so that the jet can be launched centrifugally so that the hoop stress 
associated with toroidal field may be largely responsible for the collimation.
A third possiblity, that has been discussed, is that the magnetic field be
mostly toroidal near the disk and wound up like a coiled spring
so that it can push the gas away vertically. 
Observations of protostellar outflows have as good a chance as those 
of any jets of measuring the magnetic structure. Ultimately the flow must be confined laterally 
at large cylindrical radius. It is not clear whether this is achieved by the ram pressure
of infalling gas or through the application of a quasi-static thermal pressure.
\item What is the dynamical structure of Herbig-Haro objects
and what can they tell us about the explosions that cause them? 
Forward-reverse shocks arise naturally if the velocity with which the jet is launched
varies so that the faster moving gas overtakes the slower outflow and forms a shock. The
pressure behind this shock front may be sufficient to form a reverse shock giving a 
characteristic dynamical structure. The high pressure inter-shock gas will expand
transversely, weakening the shock strengths. Again, observations should
help us to understand what is really happening.
\end{itemize}
\section*{Novae}
Classical novae, by contrast, are thermonuclear explosions which arise when hydrogen-rich gas
from a companion accumulates on the surface of a C-O or O-Ne-Mg
white dwarf and then detonates, initially uncontrollably, under degenerate
conditions. The energy release per nucleon is enough to heat the gas above the 
Fermi temperature, causing it to expand, and then above the escape energy. As several 
of the nuclear reactions involve weak interactions that take place on timescales that
are long compared with the dynamical timescales, the ejected gas is believed to 
contain many prominent radioactive species that can act as monitors of stellar activity.

X-ray novae involve similar processes occuring on the surface of a neutron star. Here, the
reactions occur much faster but the gravitational potential well is so deep that the 
gas cannot escape using its own thermal pressure. (It may be expelled by radiation pressure,
however.) Naturally, this burning will not occur uniformly over the surface of the 
star and rotational modulation of the X-ray emission was predicted and is observed.
(Curiously, the rotational frequency is observed to vary slightly, which may be due
to elevation of the X-ray photosphere by radiation pressure with approximate conservation
of angular momentum.)

My personal question list for novae is:
\begin{itemize}
\item What can be learned by observing radioactive nuclei and positron annihilation
from classical novae?
Novae are prime targets for missions like INTEGRAL and HESSI that promise to open
up the new field of MeV spectroscopy. We need to go beyond mere detection
of radioactive nuclei and use measurements of line strengths and widths to learn 
about the underlying explosion.  
\item What is the status of the beat-frequency model of QPOs?
This posited that the neutron star was rotating with a period 
similar to that of the inner regions of the accretion disk and that the observed, varying 
frequencies were a beat rather than a fundamental. The first part of this 
hypothesis has been vindicated, but I wonder about the evidence for the second part?
\item Are {\it any} QPO modes due to neutron star osciallations? 
The problem here is that some of the modes that
had been attributed to neutron star oscillations are also found in black hole systems.
(Neutron star modes can only provide the clock because the energies associated with them
are necessarily quite small.)
\item Can we measure the neutron star mass-radius relation? One of the best ways
for high energy astrophysics to repay its immense debt to nuclear physics is to measure
the equation of state of cold nuclear matter (in contrast to the hot nuclear matter
that will be explored by heavy ion colliders). This may be possible through measuring 
the gravitational redshift of atomic and nuclear lines from the surface of neutron
stars, however it is not clear what it will take to do this in practice.
\end{itemize}
\section*{Supernovae}
Supernovae are once again at center stage. In cosmology, Type Ia explosions have been modeled
empirically as one parameter standard candles, and if this is the case,
they suggest that the universe is entering a (second?) epoch of 
inflationary expansion. This is a remarkable discovery, if true.
In addition, there is circumstantial evidence that at least some types of $\gamma$-ray bursts
are associated with supernovae, both through the suggested identification 
of GRBs with star forming regions and the possible discovery of supernova light curves in a few instances.
For both lines of research to advance,
it is imperative to develop a far better understanding of the physics and the 
astrophysics of supernova explosions.

The blast waves that result from these explosions are not always 
well-described by Sedov point explosions in uniform media. 
Even if the energy release were fairly isotropic 
(and there are several reasons for suspecting that it is not)
the external medium is likely to be anisotropic. 
The beautiful images of $\eta$ Car and SN 1987a,
the former being an accident waiting to happen and the latter being one that we
are still witnessing, explain why so many mature supernova
remnants are quite non-circular despite having essentially isobaric interiors.
These supernova remnants are excellent laboratories for 
studying particle acceleration and magnetic field amplification
at shock fronts
that provide a bridge between heliospheric studies and more energetic
phenomena associated with AGN and GRBs. Non-relativistic shocks behave 
quite differently from relativistic shocks and so it is fortunate that we have 
plerions like the Crab Nebula and classic remnants like Tycho so close to home to
study.

The questions:
\begin{itemize}
\item How important are Type Ia supernova evolutionary corrections? The big concern,
as always with cosmographic studies of the expansion of the universe, is whether or not
we are confusing kinematics with physical evolution. This is particularly troubling 
here because there is no commonly agreed identity for the progenitors of these explosions
and, I believe, no consensus yet on the reason for the ``Phillips'' correction
although some promising suggestions have been made. There are
internal consistency checks and some of these have 
already been satisfied but more will 
be needed before we can sign off on the result.
\item How do we classify supernovae observationally?
I doubt that I am alone in not understanding the spectroscopic and physical distinctions
between the various types of supernova Type Ibc, Type IIn etc. I hope we can have a 
primer on the subject here.
\item What are Type Ia supernovae anyway?  Single degenerate and double degenerate models 
have their advocates. Likewise for detonation of Chandrasekhar mass
CO white dwarf versus an off center explosion in a lighter star with a helium envelope.
\item When do Type II supernovae form black holes as 
opposed to neutron stars and what are the associated rates? 
This question is timely because Chandra has just discovered a point
source inside Cas A, and as of now, the odds are about evens for it being a black hole
or a neutron star.
\end{itemize}
\section*{Hypernovae}
Gamma ray bursts continue to amaze. There has been direct verification 
that the long duration bursts are located at cosmological distances 
through the measurement of redshifts. (The same is probably true for the 
short duration bursts, although HETE2 is probably going to be necessary to 
verify this.) This leads to an impressively broad range of isotropic burst energies,
from $\sim10^{-6}$~M$_\odot c^2$ in the case of GRB 980425 to $\sim2$~M$_\odot c^2$ 
for GRB 990123 - hardly standard candles (though this has not prevented some
from trying to use them for cosmography). GRBs are now widely interpreted as 
optically thick fireballs created with large entropies per baryon, like the 
universe itself. The actual $\gamma$-ray emission, lasting for up to a few minutes,
is commonly thought to be produced by internal 
shocks in the expanding ejecta and this accounts for the great heterogeneity 
in observed $\gamma$-ray burst time 
profiles. The ninth magnitude optical burst, seen by
ROTSE from GRB 990123 (with an isotropic energy $\sim10^{-3}$ of the total)
may be caused by a reverse shock. Studying the afterglows is proving to be 
interesting in its own right, for what it has to say about the behavior of 
relativistic shocks, as a probe of the environment in which the burst occurs,
as a measure of the explosion energy and as an indicator of beaming. Broken power law spectra
are observed and these have been variously interpreted as being due
to cut-offs in the electron distribution function, radiative cooling and self-absorption.

A recent development is the circumstantial evidence for the association
of GRBs 970228, 980326, 980425 with supernovae, albeit of different types.
If the association is also with young stars, then GRBs will be invaluable 
probes of the early universe and galaxy formation. Another somewhat more secure story is
that soft $\gamma$-ray repeaters are ``magnetars''. That is to say,
their outbursts are magnetically powered and originate 
on the surfaces of young neutron stars with surface magnetic
fields $B\gtrsim10^{14}$~G.

It is hard to limit the number of questions in this subject.
\begin{itemize}
\item Are GRBs beamed? In my view, although the jet hypothesis
is eminently reasonable and fits in with some source models, 
especially collapsars, we are really only interpreting 
occasional steepening in the light curves in this manner, 
rather than seeing the clear evidence
that was provided by VLBI in the case of AGN. The argument that the bursts 
must be beamed, otherwise they would have energies in excess of a stellar
rest mass, reminds me of a similar argument in favor of them being local!
\item Are there $\gamma$-ray quiet afterglows? These are surely a prediction
of beaming. At present the observational constraints are surprisingly poor.
\item Is magnetic field amplified at external shocks? We know from observations
of young supernova remnants that relativistic protons and electrons are 
accelerated at non-relativistic shock fronts and that thermal electrons
are transmitted with temperatures below the equipartition value. 
However, in a source like 
Cas A, it appears that the magnetic field only becomes strong in the
interaction zone between the shocked interstellar
medium and the explosion debris. (It is noteworthy that even as 
impressive a radio source as Cas A is four orders of magnitude
under-luminous relative to a homogenous, maximally emitting synchrotron 
source with the same total pressure.) This can make a big difference.
When Chris McKee and I computed the nonthermal emission that would be observed
from decelarating, relativistic blast waves, we assumed that the magnetic
field is just compressed along with the gas in passing through the relativistic 
shock front. In this case, $\epsilon_{{\rm mag}}$, the ratio of the magnetic to total energy 
density is only $2v_A^2/c^2\sim10^{-9}$ in the interstellar medium, 
where $v_A$ is the Alfv\'en speed ahead of 
the shock front. More recent calculations, that are applied 
specifically to GRB afterglows, generally assume that $\epsilon_{{\rm
mag}}\sim10^{-2}$ which is 
necessary to fit the fluxes (although the scaling laws
are unchanged). For this reason and because the best studied 
GRB afterglow,
GRB 970508, shows no sign of a mildly relativistic transition
in the particle acceleration efficiency,
I still suspect that the afterglow emission 
originates well downstream  from the outer shock.
\item Are GRBs really associated with supernovae? The late time light curve
seen in GRB 980326 could have a different 
explanation. In particular, as Ann
Esin and I have been considering, it fits
rather well with scatttering of the initial optical burst by dust just outside
the sublimation radius. 
This occurs typically at a distance $\sim10^{18}$~cm.
As refractory dust has high albedo and is forward scattering, the characteristic
delay is plausibly a few months, as observed.
\item How easy is it to have an ultrarelativistic jet emerge from inside a collapsing
star? Entrainment of gas might easily occur and prevent the flow from
attaining bulk Lorentz factors $\sim300$.
\item Are the ``cyclotron'' lines real and the $\sim300$~keV ``breaks'' 
generic and, if so,
can they be formed in ultrarelativistic outflows? Existing explanations
seem a little contrived.
\item What is the underlying physical mechanism for creating the fireball? 
Most models now seem to involve black holes, magnetic field and wishful 
thinking. The problem is hard. The main challenge is to 
amplify the magnetic field fast enough to make an electromagnetic bomb.
Fortunately there are new ingredients in the strongly curved
spacetime around a black hole or a pair of orbiting neutron stars.
The orbits can precess differentially at near relativistic speed and
this can lead to an extremely rapid field growth - faster than 
conventional dynamos and, indeed, faster than exponential.  This,
in turn, induces $\gtrsim10^{22}$~V EMFs which cannot be shorted out and accelerate pairs directly. 
\end{itemize}
\section*{Jets}
The black hole model of AGN has been vindicated observationally,
and, beyond all reasonable doubt, most normal galaxies, like our own 
contain central black holes with masses in the million to billion solar mass
range. There are also at least nine well measured compact object masses in 
excess of $\sim2.5$~M$_\odot$ that are surely also black holes. 
There are promising, but so far, less compelling indications that  
some of these holes spin rapidly.

Jet are a common, though not universal, accompaniment of accretion which 
suggests that they are involved in carrying off some of the energy
and angular momentum released by the infalling gas.
However, we do not understand how they are formed or even if there is a 
universal mechanism at work. We do know that magnetic field must grow 
to dynamically significant levels in accretion disks and most jet formation 
models now involve magnetic field except perhaps when the mass accretion 
rate greatly exceeds the Eddington rate. 

Our understanding of the emission from jets has also advanced. 
The discovery of rapidly variable GeV and TeV $\gamma$-rays from blazars
shows that they can be extremely luminous. Radio astronomers have been 
mapping the smoke not the fire. The standard radio synchrotron model
is also under assault. There is increasing evidence 
that compact components have brightness temperatures well in excess of the inverse 
Compton limit, even allowing for plausible Lorentz factors. Large degrees of
circular polarization are also being reported. All of this suggests
that some alternative, possibly coherent emission mechanism is at work
at least in the compact cores. 
(Note, that it is not sufficient to explain how high brightness radio emission
is {\it emitted}. It is also necessary to explain how it is {\it transmitted}
out of the nucleus when there are many potential non-linear scattering 
mechanisms which will degrade the brightness temperature.) 

In an impressive display of the power of VLBI, the radio astronomers
have been able to show that the M87 jet is collimated within $\sim 100m$.
If relativistic jets are powered by the black hole itself or the gas flow around the black hole
then their energy has to be carried {\it initially} in some form
other than electron-positron pairs, which are subject to catastrophic radiative losses.
Electromagnetic Poynting flux is the prime suspect. In other words,
relativistic jets are starting to resemble pulsars. (Contrariwise,
the Crab pulsar now appears to form a pair of ``jets'').

My final list of questions is: 
\begin{itemize}
\item Are AGN jets hyper-relativistic? Radio jets exhibit bulk Lorentz factors
$\Gamma\sim10$; $\gamma$-ray burst models have taken us over the psychological hurdle
to $\Gamma\sim300$ which might just account for the reported radio
variability of jets under 
the synchrotron model were it not to imply {\it steady} $\gamma$-ray burst level powers in AGN
jets.
Hence the appeal to coherent processes. Before we solve this problem, though, we must 
identify the jet working substance and if, and where, Poynting flux is transformed
to plasma. (This last is still an interesting question in the case of the Crab pulsar wind.)
\item Are jets better approximated as episodic or steady? Traditionally, we have modeled
jets as stationary flows upon which have been imprinted perturbative disturbances
which form shock fronts - the emitting elements. However, GRS 1915+105 suggest a quite different
model - jets as a sequence of small explosions, perhaps associated with intermittent flow
in the accretion disk, that expand into and keep open an evacuated channel. (YSO jets
offer support to both views.)
\item How are jets collimated? Ordered and disordered, poloidal and toroidal
field have all been proposed for launching and collimating  AGN jets from disks,
just as with the YSO jets.
3D MHD global simulations are becoming increasingly ambitious and ever more
relevant.
\item Are relativistic jets powered by the spin energy of the hole or the binding energy 
of the accreting gas? The former seems more likely to form ultrarelativistic outflows;
the latter may, on average, release more power.  General relativistic numerical
simulations are starting to guide our intuition.
\item Why are there no gamma-ray megabursts? GRBs are thought to be 
associated with the birth of 
stellar black holes and produce powers of up to $\sim10^{-7}c^5/G$ for $\sim10^6m$.
If massive black holes are formed with masses $\sim10^6$~M$_\odot$ at a rate 
of several per year, we might expect
to see megabursts with similar powers but lasting for months. We don't. Perhaps, instead,
massive black holes grow from much smaller holes, which, themselves,
might be relics of the first generation of stars which may have masses $\sim10^3-10^4$~M$_\odot$
at $z\sim30$. 
\item How much do AGN contribute to the luminosity density of the universe? 
This is a closely related question. The measurement of the far infared background,
the discovery of hard X-ray emission from some Seyfert galaxies and the spectrum of the 
X-ray background all point to AGN power being a significant fraction
of the stellar luminosity density. If so, then there are probably implications
for galaxy formation and development. For example, elliptical galaxies
may result when a black hole grows rapidly and early in the life of the galaxy
so that it is capable of blowing away late infalling gas before it can form 
a disk.
\end{itemize}
\section*{Connections}
As I hope this brief introduction has brought out, there are strong inter-connections 
between our studies of these different types of cosmic explosion. Accretion disk
coronae can look quite like their solar counterpart. Novae have some  dynamical
similarities to miniature supernovae whose remnants, in turn, behave quite like aging
$\gamma$-ray burst afterglows. Similar electromagnetic processes are at work
around pulsars and black holes. $\gamma$-ray bursts themselves have some 
similarities, at least radiatively, with the early universe. And so on. 
\section*{Acknowledgements}
I acknowledge support under NASA grant 5-2837 and NSF grant 
AST 99-00866 and Re'em Sari for comments.
\end{document}